\documentstyle[aps,prb,epsfig,multicol]{revtex}
 \begin{document}
\draft
\title{Field evolution of vortex lattice  in $\rm
LuNi_2B_2C$ seen by decoration in fields up to 1.5 kOe}
\author{L.Ya.~Vinnikov, T.L.~Barkov }
\address{Institute of Solid State Physics, Chernogolovka,
Moscow Region, 142432 Russia}
\author{P.C.~Canfield, S.L. Bud'ko, and V.G.~Kogan}  
\address{Ames Laboratory DOE and Department of Physics, 
Iowa State University, Ames, Iowa 50011, USA}
\maketitle

\begin{abstract}
Evolution of the vortex lattice, the rhombus-square
transition included, for $\rm LuNi_2B_2C$ in the field along the $c$
crystal axis, is tracked by the decoration technique pushed up
to the record high (for this method) field of 1480 Oe. Decoration
images are analyzed with the help of the Fourier transform, which
indicates disordered structures in small fields of a few Oe. In fields 
$H<200\,$Oe the coexisting domains of different structures are
observed. A technique based on the Fourier transform is employed
 to see the domains separately. The transition
to the square lattice is recorded near 900 Oe. The  results are in
agreement with predictions of the nonlocal London theory.
\end{abstract}


 \begin{multicols}{2}
\narrowtext

\section{Introduction}
  The family of the rare-earth ($R$) metallic nickel-borocarbides
$\rm RNi_2B_2C$ has been the subject of considerable  recent attention 
due to the interplay of  magnetic and superconducting
properties.\cite{Hilscher,Canfield} The critical temperature 
$T_c$ of these materials depends on the particular rare earth
(or their combination), or on material purity, and ranges from
zero to about $16\,$K in nonmagnetic members such as $\rm
LuNi_2B_2C$. The vortex  lattices (VL) in
these compounds exhibit a variety of phases with changing magnetic
field and temperature. Studies of VL's are facilitated by
availability of large high quality single crystals\cite{Canfield} with
a relatively large Ginzburg-Landau parameter, $\kappa\sim 10-20$, low
pinning, and a broad region of the $HT$ phase diagram where the
superconducting magnetic properties are nearly reversible. 

The structural phase transition from the triangular to 
square VL driven by the magnetic field is of a particular  
interest. A variety of experimental  techniques are used 
to study this transition:   small-angle neutron scattering
(SANS)\cite{Eskildsen,Yethiraj,McPaul},   scanning tunneling
microscopy (STM)\cite{Y.De Wilde}, and   Bitter decoration
by small magnetic particles.\cite{Eskildsen,Abrahamsen} The SANS and
STM  methods  require rather high magnetic fields (few
kOe and higher), whereas  decoration is commonly employed for fields
on the order of several hundred Oe or less because of a relatively low
resolution.

For $\rm ErNi_2B_2C$, the phase transition from the triangular to
square VL in increasing field has been recorded by SANS and
complemented by decoration.\cite{Eskildsen} Decoration
patterns revealed coexistence of the square and triangular lattices in
fields near 600 Oe. In $\rm LuNi_2B_2C$  crystals, the square vortex
lattice for  high magnetic fields of a few Tesla  has been seen in
STM.\cite{Y.De Wilde} No experiments in low magnetic fields were
carried out for this material. The theoretical model, which describes
most of the features of evolving VL's is based on the London theory
corrected for nonlocal effects.\cite{Kogan}  
 
In this work the  VL's in $\rm LuNi_2B_2C$ were studied by the 
decoration method. To cover the field range of greatest interest  for
the VL evolution, the method was extended up to fields of  
1.5 kOe, the field region unprecedented for the decoration
technique in materials with $\kappa\gg 1$. To our knowledge, the
maximum field at which the decoration has been attempted was $1100$ Oe
on Nb.\cite{Brandt_Ess}

 \section{Experimental}

  We have used single crystals $\rm LuNi_2B_2C$  of the size
  $\sim 3\times$5 mm in the $ab$-plane and $\sim 
 0.5$ mm thick. The crystals were grown from ${\rm Ni_2B}$ flux,
 as described elsewhere.\cite{Canfield,cho} Experiments were carried
out in the field cooled regime with the field nominally parallel to
the $c$ axis (within a few degrees). For all fields except $1480\,$Oe, 
the temperature was set  4.2 K initially  but may have  risen  to 7-9 K
during the decoration experiment. For $1480\,$Oe, the initial
temperature was 1.5 K. To minimize effects of inhomogeneities and
pinning,  decorations were done on a particular location of the same
sample which was cleared  by acetone after each decoration. A scanning
electron microscope
 was used to resolve decoration patterns on the sample surface and to
determine  VL orientations  relative to the crystal by comparing
decorations with electron channeling patterns  (ECP). 
 
  Routinely,  the decoration experiments  are done in  
  fields  near the lower critical field $H_{c1}$ on samples as
thin platelets in the perpendicular magnetic field.  Since there is no
quantitative theory   to estimate
the resolution limit, we use the following qualitative  guidelines 
  in pushing the method to higher fields:

 1. The size of magnetic particles should be  small relative
to  the London penetration depth $\lambda(T)$ and to the 
intervortex spacing $\sim \sqrt{\phi_0/B}$.
 
2. The density of particles should be  small to avoid clustering. 
  
3. The field gradients due to VL's should be large enough 
for the interaction with the particle magnetic moment $\mu$ to exceed
the thermal energy, $\mu H > 3k_BT$.\cite{Trauble}
 
      The first condition is realized by manipulation of
the buffer helium pressure within $\approx 0.1\,$torr.
The optimal size of particles of $5-10\,$nm should not
limit the resolution up to fields $\sim  10\,$kOe. The second
requirement is met by empirically minimizing the quantity of the
evaporated magnetic material and by varying the distance between the 
heater and the sample. The most difficult to control and to satisfy 
is the last condition because the magnetic moment $\mu$ of iron
particles  is unknown. Besides,  the actual temperature of decoration
is  uncertain because of the heating during the iron
evaporation.  Still, one has to keep this $T$ as low as
possible. To achieve this, we used onset temperatures $1.5-2\,$K for
decoration in high fields. In this work, we have found that the
contrast images of decorated VL's in  $\rm LuNi_2B_2C$ can be readily 
resolved in fields up  to 1480 Oe.
 
The information one can extract from decoration experiments in 
fields  below $H_{c1}$ differs from the case $H \geq H_{c1}$. 
 In the first case, the intervortex distance exceeds the
penetration depth $\lambda$, and one can well resolve
single vortices. However, the intervortex interaction in this situation
is too weak to arrange vortices in a well ordered lattice, given the
competition with disordering pinning. Nevertheless, one can utilize the
decoration data to obtain a rough estimate of the penetration depth. 
 The field distribution of a single vortex in the bulk has a
characteristic diameter of 2$\lambda$.
  Near the sample surface the vortex field ``opens up".\cite
{Pearl} As a result, the field gradients at the sample surface are
lower than in the bulk. The field lines at a distance 
 $\lambda$ from the vortex core in the bulk, cross the
surface at a larger distance from the core. 
 One may consider the magnetic diameter of the vortex at the
surface as  $L\approx m\lambda$, where the empirical expansion
factor $m$ can be determined measuring the size $L$ of the iron
particle clusters forming at each vortex site during the low
field decoration.   Data show that $m\approx 4$ provides a
good estimate of $\lambda$. 

In applied fields well above $H_{c1}$, the fields of 
neighboring vortices overlap, and the resolution of decoration
experiments drops.  Nevertheless,  the  contrast of the decoration
patterns still  provides enough quantitative information on the VL
symmetry and intervortex spacing.  The computer
processing of the images (fast Fourier transform and image
filtering followed by inverse Fourier transform)  greatly
enhance our ability to study VL's and the domain
structures in fields which were thought inaccessible to the
decoration technique. In this work,  VL's   
were successfully decorated and analyzed in a broad field
interval from 4 to 1480 Oe.

\section{Results}

\begin{figure}
\epsfxsize= .8\hsize
\centerline{
\vbox{
\epsffile{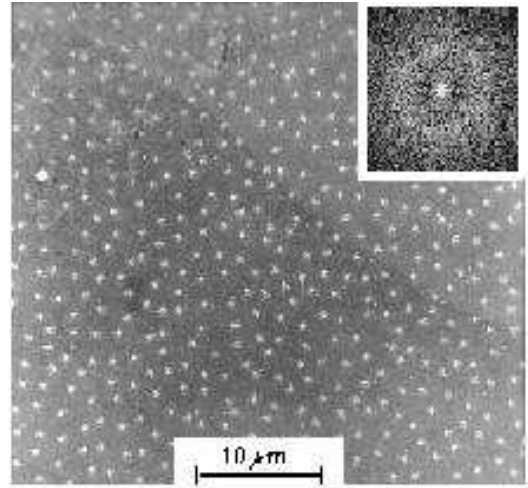}
}}
\vskip \baselineskip
\caption{  Vortices  in field $H=4.5\,$Oe
  along the $c$ crystal axis.  The Fourier transform in the inset  
indicates that the vortex system is an amorphous solid.}
\label{4oe}
\end{figure}

As is seen in Fig. \ref{4oe}, in small fields vortices do not
form a well ordered lattice.  A straightforward count shows
that the average intervortex spacing corresponds to one  
flux quantum per vortex. 
 
 High magnification images of single vortices reveal the
average size of the iron clusters of about 600 nm. Taking this
as an estimate of $4\lambda$   at  $T\approx
0.5\,T_c$, we obtain $\lambda_0\approx   130 \,$nm utilizing the
two-fluid approximation $\lambda^2(T)=\lambda_0^2/(1-T^4/T_c^4)$. This
value is larger than $71\,$nm obtained from the $H_{c1}$
data,\cite{Takagi} and $ \approx 100 \,$nm (Ref.
\onlinecite{Budko})  obtained from the magnetization measurements, but
close to the estimate of Ref. \onlinecite{Hilscher}. 

 \begin{figure}
\epsfxsize= .8\hsize
\centerline{
\vbox{
\epsffile{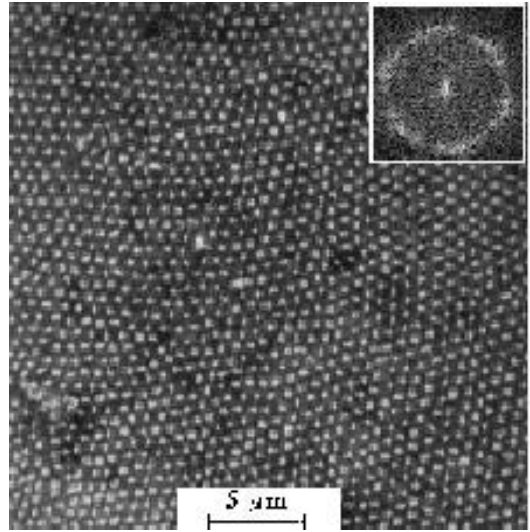}
}}
\vskip \baselineskip
\caption{Vortex system for $H=40$ Oe. Domains of well ordered lattices
are seen. The maxima of the Fourier transform are smeared due to
varying orientations of the domains.}
\label{40oe}
\end{figure}
 In higher fields, intervortex interaction  is stronger and vortices form 
ordered lattices. Starting from $\approx$40 Oe and up to $\approx$200 Oe we
observe domains of nearly regular triangular  VL's; see Fig. \ref{40oe}.
VL  orientations within each domain are quite random.  Typical size of  a
domain is  a few intervortex distances and increases with the magnetic
field.

\begin{figure}
\epsfxsize= .8\hsize
\centerline{
\vbox{
\epsffile{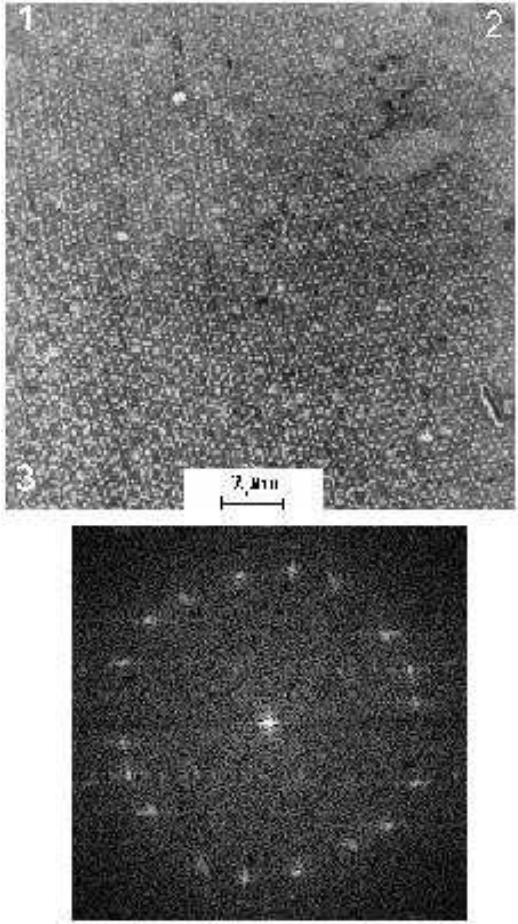}
}}
\vskip \baselineskip
 \caption{The upper panel: the three-domain structure for  $H=
200\,$Oe. The domains (marked as 1,2,3) are shown  separately in Figs.
\ref{dom3}. The lower panel: Fourier transform  of the   upper
panel pattern indicates the presence of three well ordered domains.
Each of the domains is responsible for 6 maxima forming nearly a
hexagon. 
\label{domains}  }
\end{figure}

\begin{figure}
\epsfxsize= .8\hsize
\centerline{
\vbox{
\epsffile{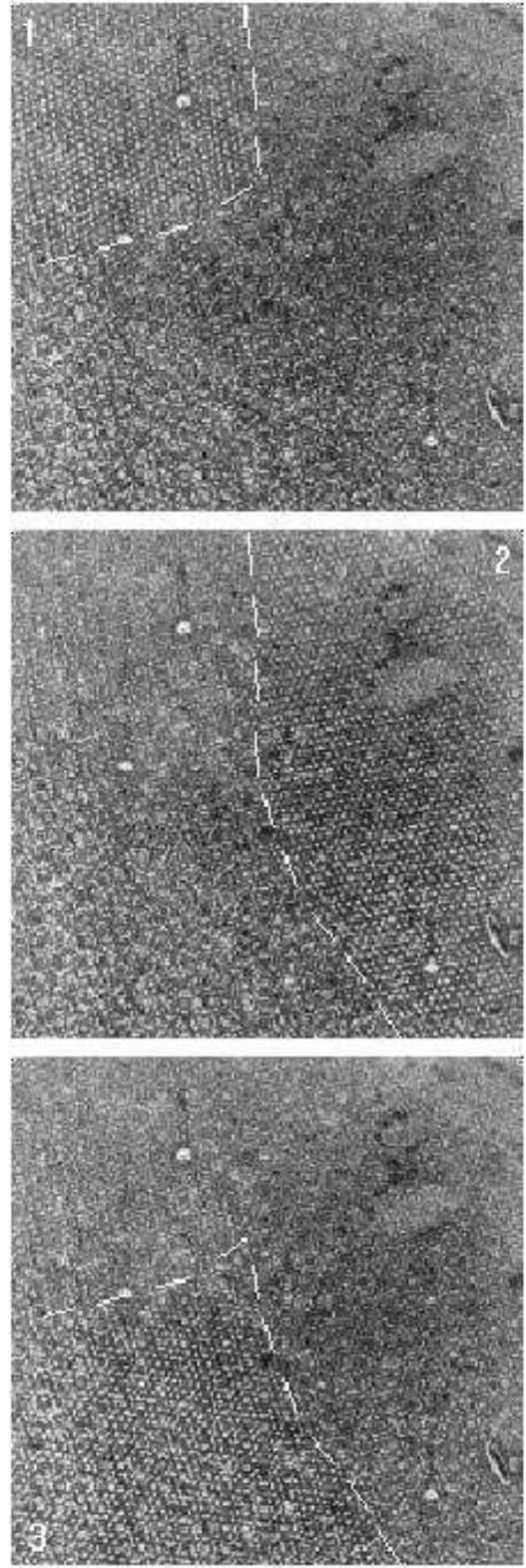}
}}
\vskip \baselineskip
 \caption{\label{dom3}The three panels show three different VL's. Each
one is obtained by the inverse Fourier transform of a selected sets of
six maxima out of 18 of Fig. \ref{domains}.  }
\end{figure}

 An example of VL at $H=200\,$Oe is shown in the upper panel of Fig.
\ref{domains}. The Fourier transform (FT) of this pattern shows
that the structure consists of three nearly regular triangular domains
with different VL orientations.  
 By choosing the FT maxima which belong to one of the domains and
clearing all other reflexes one can obtain with the help of the
inverse FT the patterns of all three domains in real space
separately. These patterns are shown in  Fig. \ref{dom3}.
 Note that the domain size (in units of intervortex distance) for 200
Oe is large as compared to domains in the field of 40 Oe.  

\begin{figure}
\epsfxsize= .9\hsize
\centerline{
\vbox{
\epsffile{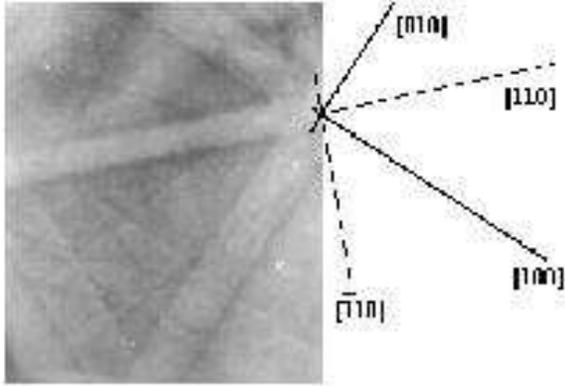}
}}
\vskip \baselineskip
\caption{Electron channeling pattern   for the same
place as   the decoration of Fig. \ref{domains}. \label{ecp} }
\end{figure}

The influence of crystal inhomogeneities and anisotropy on the VL domain
structure and ordering has been extensively studied for ${\rm 
NbSe_2}$.\cite{Bolle,Pardo} The nonlocality brings a new complexity to the
process of domain formation because VL's now depend on $T,H$ and the local 
mean-free path.\cite{ell} Comparing the ECP picture of Fig. \ref{ecp} with
Figs. \ref{dom3}, we find that one of the closepacked directions on the
upper panel is aligned with $\langle{\bar 1}10\rangle$ direction of
the crystal, whereas that of the middle panel is aligned with $\langle
110\rangle$. It is worth noting that the model of Ref.
\onlinecite{Kogan} predicts that the short diagonal of the rhombic VL
cell in small fields  should point in one of these directions.

\begin{figure}
\epsfxsize= .8\hsize
\centerline{
\vbox{
\epsffile{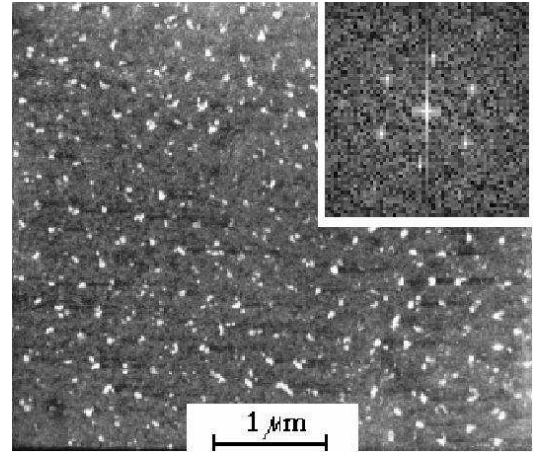}
}}
\vskip \baselineskip

\epsfxsize= .8\hsize
\centerline{
\vbox{
\epsffile{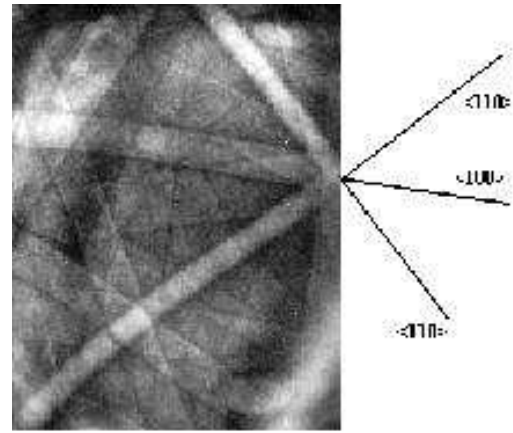}
}}
\vskip \baselineskip
\caption{ 
 The upper panel shows the decoration pattern for $H=300$ Oe with the
Fourier transform in the inset. The lower panel  shows the
 corresponding electron channeling pattern indicating that one of the
VL close-packed directions coincides with $\langle 100\rangle$ crystal
direction.\label{300oe} }
\end{figure}

  In fields more then 200 Oe, the typical domain size considerably
exceeds  the intervortex distance. The orientation of the
closepacked VL direction in most domains correlates (within few
degrees) with  $\langle 100 \rangle$ crystallographic direction;
this is seen in Fig. \ref{300oe} in which the decoration pattern for
$H=300\,$Oe  and the ECP picture are provided.

\begin{figure}
\epsfxsize= .8\hsize
\centerline{
\vbox{
\epsffile{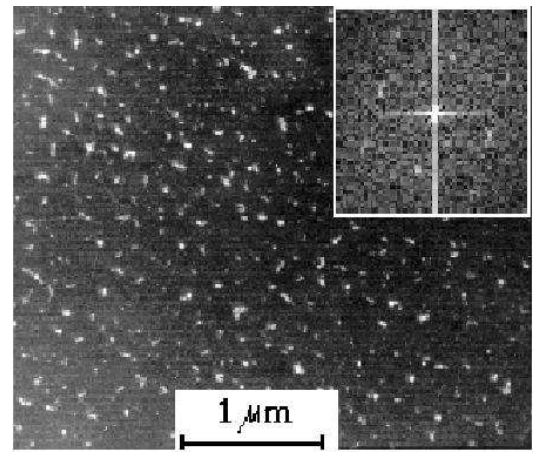}
}}
\vskip \baselineskip
\caption{
 The applied field is $H=710$ Oe.  The direction analysis of the
Fourier transform  in the inset gives $\beta\approx 82^{\circ}$ and the
corresponding diagonal of the rhombic cell aligned with $\langle
100\rangle$.\label{710oe}  }
\end{figure}

Still, in other domains the closepacked  direction coincides with
$\langle$110$\rangle$, the fraction of which decreases with field 
and  at $H=600\,$Oe  reaches $\approx 1/3$. The VL unit cell is nearly
rhombic with the apex angle $\beta$  exceeding $60^{\circ}$, as can be
extracted from the FT in  Fig. \ref{710oe}. The angle $\beta$
increases  with field. The preferable orientation of the  VL
corresponds to the unit cell diagonal opposite to $\beta$  being
aligned with $\langle$100$\rangle$.

 In fields of about 600 Oe and higher, in some surface locations we
observe domains of the square VL, whereas majority of the
domains contain strongly distorted triangular
(or, better to say, rhombic) VL, see FT in Fig. \ref{710oe}. 
The fraction of domains with the square lattice increases with
increasing field. A similar situation has been seen in SANS data on
$\rm V_3Si$, where the presence of domains was deduced from analysis
of the scattering maxima.\cite{Mona} Here, we see the domains
directly.   
 
\begin{figure}
\epsfxsize= .85\hsize
\centerline{
\vbox{
\epsffile{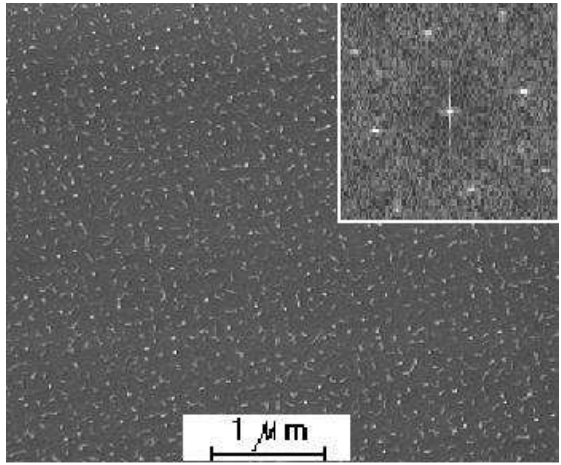}
}}
\vskip \baselineskip
\caption{
The vortex lattice in the $\rm LuNi_2B_2C$ single crystal for 
the field  $1480\,$Oe parallel to the $c$ axis. 
 \label{1480oe}  }
\end{figure}

The square domains are dominant in fields of about 1000 Oe and higher, 
Fig. \ref{1480oe}. The square diagonals are parallel to
$\langle 100\rangle$ and $\langle 010\rangle$ crystallographic
direction. 

\begin{figure}[t]
\centerline{
\epsfxsize=0.9\columnwidth
\epsffile{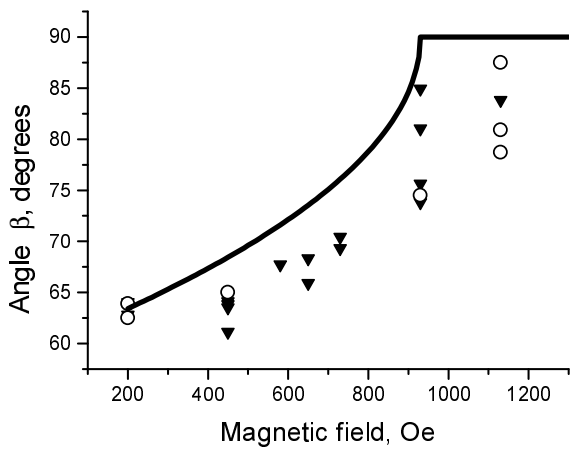}
}
\caption{The apex angle $\beta$  versus $H$. The triangles and the
circles are obtained at two different locations on the sample
surface. A few triangles (circles) for the same field
indicate that the data points are extracted from a few
different  locations of the same decoration patch.\label{ngraph}}
\end{figure}

The results of our study of VL's in $\rm LuNi_2B_2C$ for a two
particular locations on the crystal surface are collected in Fig.
\ref{ngraph} where the  field dependence of the angle $\beta$  is
shown. The triangles and circles show the decoration results from two
different locations of the crystal. Few points for the same field
indicate the scattering of the angle $\beta$  within the decoration
area of about $100\times 100\,\mu$m. In the field interval 200-600 Oe,
the apex angle $\beta$ of the rhombic unit cell  slowly increases with
field.  With the further field increase, VL's deviate even faster from
the standard triangular VL (with $\beta=60^{\circ}$) toward larger
apex angles ($\beta$ exceeds $70^{\circ}$) and in many decorations
coexistence of domains with triangular and nearly square VL's are 
observed. 

\section{Discussion}

 The theory of Ref. \onlinecite{Kogan} predicts that in fields less
than $H_1\sim 200\,$Oe along $\langle 001\rangle$ crystal direction,  the
VL should have a rhombic unit cell with a short diagonal along either
$\langle{\bar 1}10\rangle$ or $\langle 110\rangle$.
The apex angle $\beta$ of the rhombus should be less than $60^{\circ}$. 
Experimentally, up to fields about 200 Oe we did not record a well
ordered VL.  Instead, a structure made of many domains with different
lattice orientations is observed. The prediction for the value of
$\beta <60^{\circ}$ is also hard to verify in this field region.

One can argue that in small fields energy differences 
between different possible VL configurations are
comparable with characteristic pinning energies.  In this situation, the
VL structure is  affected by distribution of pinning sites  which
may vary throughout the sample.  Therefore, we need better crystals of
 $\rm LuNi_2B_2C$ to support or disprove the low field theoretical
scenario for equilibrium VL's for this material (for  $\rm YNi_2B_2C$
this scenario has been confirmed\cite{McPaul} by SANS data in fields up
to 1200 Oe). 

Still, the low field decoration results can be used 
 to estimate the London penetration depth $\lambda$ at decoration 
temperatures of  $(0.3 - 0.5) T_c$. The  value of
$\lambda$ obtained by measuring sizes of the particle clusters, is in a
 reasonable agreement with values obtained by other
techniques.\cite{Hilscher,Budko}
 It is to be noted that  $\lambda$ estimated by measuring the
"vortex diameter" $d$ in decoration experiments  is in a  fair 
agreement with other data also for YBCO,\cite{Vinnikov88,YBCO} for
which an empiric relation $d=4\lambda$ yields $\lambda_0\approx
150\,$nm and for NbSe$_2$ for which the procedure gives $\approx
270\,$nm.\cite{march} 

 With the magnetic field increasing, the role of pinning diminishes
because the intervortex interactions become stronger. The field region
with $H$ exceeding approximately 300 Oe is more amenable for checking
the theoretical model of Ref. \onlinecite{Kogan}.  The model is based
on the nonlocal London approach and as such requires a number
of microscopic input parameters (the Fermi surface, the penetration
depth, the mean-free path, to name a few) for a quantitative
comparison with experiment. Here, we use a simpler ``mean-field"
approach based on the assumption that the structural transition from
the rhombic to the square VL at the transition field we call $H_2$ is
of the second order.  This assumption is supported by numerical
simulations of Ref. \onlinecite{Kogan} which show that the
rhombus-to-square  transition  at $H_2$ is accomplished by an
infinitesimal deformation so that at the transition the structure
changes continuously. Moreover, the square phase has an extra symmetry
element: the 4-fold rotation axis which is absent in the rhombic cell.
Both of these features are characteristic of a 2nd order transition.
We, therefore,  can apply the general Landau
theory according to which the energy density near the transition at
$H_2$  can be written as 
\begin{equation}
F\{\eta;B\}=F_0(B) + a (B-H_2)\eta^2 + b\eta^4/2\,.
\label{F}
\end{equation}
Here, $F_0$ is the part unrelated to the VL structure; the magnetic
induction  $B$  for platelet-like samples is equal to the perpendicular
applied field $H$. The field independent and positive
 coefficients $a,b$ are related to a particular VL; $\eta$ is a properly
defined order parameter which should be zero in equilibrium above $H_2$. 
Since the apex angle of the high-field structure is $\pi/2$, one can 
choose $\eta=\pi/2-\beta$. Note that the field $B$ in (\ref{F}) plays
the role of temperature in the standard Landau theory.  

The equilibrium structure corresponds to
 $\partial F/\partial \eta= 0$ which yields $\eta=0$ above $H_2$ and 
  \begin{equation}
 \beta^{(1,2)} = {\pi\over 2} \mp\sqrt {{aH_2\over b}\Big (1-{B\over
H_2}\Big)}\,.
\label{beta}
\end{equation}
for $B<H_2$. The two signs correspond to two possible structures with apex
angles $\beta^{(1)}+\beta^{(2)}=\pi$;  these
are rhombic VL's rotated relative to each other by $\pi/2$.   We then 
expect $(\pi/2-\beta)\propto\sqrt{H_2-B}$ near $H_2$. In fact, the
numerical simulations show that with good accuracy this dependence
holds all the way down to small fields where $\beta$ should approach
$\pi/3$, i.e., $\sqrt{aH_2/b}\approx \pi/6$.  Thus, the apex angle
$\beta$ (in degrees) should satisfy an approximate relation
  \begin{eqnarray}
 \beta  &\approx& 90^{\circ} -30^{\circ}\sqrt { 1-B/H_2}\,,\quad
B<H_2\,,\label{beta1}\\
\beta&=&90^{\circ}\,,\qquad\qquad\qquad\qquad\quad B>H_2\,\nonumber
\end{eqnarray}
in a broad field domain near $H_2$. This formula is convenient for
comparison with data since it contains only one fitting parameter
$H_2$. The solid line in Fig. \ref{ngraph} is calculated with the help
of Eq. (\ref{beta1}) with $H_2=930$ Oe.
 
There are few reasons for a considerable spread of $\beta$
values shown in Fig. \ref{ngraph}.  Since the lattice should undergo
restructuring during the cooling, the cooling rate may affect the
resulting structure. The gas pressure may influence the actual
decoration temperature since during the ``evaporation" of iron
particles the sample heats up and cools down with the pressure
dependent rate. Clearly, the spread related to different loci is
related to crystal inhomogeneities. Moreover, the nonlocal London
model predicts\cite{Kogan} and the data\cite{ell}  confirm  that the
transition field $H_2$ is sensitive to the crystal purity ({\it via}
the mean-free path $\ell$). Even weak inhomogeneities in  $\ell$
within one decoration patch of 
$100\times 100\,\mu$m$^2$ may influence VL structures.  As is seen
from Eq. (\ref{beta1}), $d\beta/dB$ diverges as $B\rightarrow H_2$,
in other words, a small spread in values of $H_2$ may cause a
considerable variation in $\beta$. This statement can also be
formulated as enhanced "structural fluctuations" near the second
order transition at $H_2$. In fact, our data show that the $\beta$
spread peaks in the field interval $700 - 1000\,$Oe, i.e., just near
$H_2$.  

\acknowledgements
We thank N.S. Stepakov, L.G. Isaeva, D.K. Finnemore, and J.E. Ostenson
for assisting in this work.  L.V. and T.B. acknowledge the
support of Russian State Program ``High-temperature superconductivity,
Volna 4-Vin". The work of L.V. was funded by Russian State
Contract 107-1(00)-P and in part by a COBASE grant of the National
Research Council. Ames Laboratory is operated for US DOE by the Iowa
State University under Contract No. W-7405-Eng-82.

\end{multicols}
 \end{document}